\documentclass[prl,aps,superscriptaddress,subcaption,twocolumn,preprintnumbers,amssymb,nobibnotes,raggedbottom,10pt]{revtex4-1}
\usepackage{CJK,hyperref,graphicx,amsmath,mathrsfs,titlesec,tikz}
\graphicspath{{figures/}}
\titleformat{\section}{\centering\normalsize\normalfont\bf}{\thesection}{0em}{}
\hypersetup{pdftitle={},pdfcreator={},linkcolor=[rgb]{0.15,0.35,0.75},colorlinks=true,citecolor=[rgb]{0.675,0,0.2},urlcolor=[rgb]{0.15,0.35,0.65}}
\definecolor{mhvblue}{rgb}{0.6,0.6,0.7765}
\definecolor{hblue}{rgb}{0,0,0.575}
\definecolor{hred}{rgb}{0.575,0.0,0.225}
\definecolor{hgreen}{rgb}{0.0,0.4,0.2}
\definecolor{hteal}{rgb}{0.0,0.545,0.7451}
\newcommand{\eq}[1]{\vspace{-0.5pt}\begin{equation}#1\vspace{-0.5pt}\end{equation}}
\newcommand{\fwbox}[2]{\text{\makebox[#1][c]{$\hspace{-150pt}\displaystyle#2\hspace{-150pt}$}}}
\newcommand{\fwboxL}[2]{\text{\makebox[#1][l]{$#2$}}}
\newcommand{\fwboxR}[2]{\text{\makebox[#1][r]{$#2$}}}
\newcommand{\equivR}{\fwbox{14.5pt}{\hspace{-0pt}\fwboxR{0pt}{\raisebox{0.47pt}{\hspace{1.25pt}:\hspace{-4pt}}}=\fwboxL{0pt}{}}}
\newcommand{\equivL}{\fwbox{14.5pt}{\fwboxR{0pt}{}=\fwboxL{0pt}{\raisebox{0.47pt}{\hspace{-4pt}:\hspace{1.25pt}}}}}
\newcommand{\fig}[3]{\raisebox{#1}{\includegraphics[scale=#2]{#3}}}
\newcommand{\bigger}[1]{\raisebox{-0.95pt}{\scalebox{1.25}{$#1$}}}
\newcommand{\mi}{\raisebox{0.75pt}{\scalebox{0.75}{$\hspace{-0.5pt}\,-\,\hspace{-0.5pt}$}}}

\renewcommand{\phi}{\varphi}
\renewcommand{\r}[1]{{\color{hred}#1}}
\renewcommand{\b}[1]{{\color{hblue}#1}}
\newcommand{\g}[1]{{\color{hgreen}#1}}
\renewcommand{\t}[1]{{\color{hteal}#1}}

\begin{document}
\begin{CJK*}{UTF8}{}
\CJKfamily{gbsn}
\title{\texorpdfstring{The Four-Point Correlator of Planar sYM at Twelve Loops\\[-22pt]~}{The Four-Point Correlator of Planar sYM at Twelve Loops}}
\author{Jacob~L.~Bourjaily}\email{bourjaily@psu.edu}
\affiliation{Institute for Gravitation and the Cosmos, Department of Physics,\\Pennsylvania State University, University Park, PA 16802, USA}
\author{Song He (何颂)}
\email{songhe@itp.ac.cn}
\affiliation{CAS Key Laboratory of Theoretical Physics, Institute of Theoretical Physics, Chinese Academy of Sciences, Beijing 100190, China}
\affiliation{School of Fundamental Physics and Mathematical Sciences, Hangzhou Institute for Advanced Study, Hangzhou, Zhejiang 310024, China}
\author{Canxin Shi (施灿欣)}
\email{shicanxin@itp.ac.cn}
\affiliation{CAS Key Laboratory of Theoretical Physics, Institute of Theoretical Physics, Chinese Academy of Sciences, Beijing 100190, China}
\author{Yichao Tang (唐一朝)}
\email{tangyichao@itp.ac.cn}
\affiliation{CAS Key Laboratory of Theoretical Physics, Institute of Theoretical Physics, Chinese Academy of Sciences, Beijing 100190, China}
\affiliation{School of Physical Sciences, University of Chinese Academy of Sciences, No.19A Yuquan Road, Beijing 100049, China}


\begin{abstract} 
We determine the 4-point correlation function and amplitude in planar, maximally supersymmetric Yang-Mills theory to 12 loops. We find that the recently-introduced `double-triangle' rule in fact implies the previously described square and pentagon rules; and when applied to 12 loops, it fully determines the 11-loop correlator and fixes all but 3 of the ($22,\!024,\!902$) 12-loop coefficients; these remaining coefficients can be subsequently fixed using the `(single-)triangle' rule. Not only do we confirm the Catalan conjecture for anti-prism graphs, but we discover evidence for a greatly \emph{generalized Catalan conjecture} for the coefficients of all \emph{polygon-framed fishnet} graphs. We provide all contributions through 12 loops as ancillary files to this work. \\[-18pt]%
\vspace{-10pt}
\end{abstract}
\maketitle
\end{CJK*}

\vspace{-15pt}\section{Introduction}\label{introduction_section}\vspace{-14pt}
Much of the recent progress in our understanding of perturbative Quantum Field Theory has resulted from the discovery of remarkable new structures within the `theoretical data' resulting from hard computations; such discoveries have led to many deep insights and fueled the development of powerful new tools for computation---extending our theoretical reach to further discovery.

The four-point amplitude in the planar limit of maximally supersymmetric ($\mathcal{N}\!=\!4$) Yang-Mills theory (sYM) has long served as an important benchmark (among many) in our perturbative reach. This amplitude was first determined at the integrand level via generalized unitarity through six loops \mbox{\cite{Bern:1997nh,Anastasiou:2003kj,Bern:2005iz,Bern:2006ew,Bern:2007ct,Bern:2012di}}, to eight loops using the `soft-collinear bootstrap' in \mbox{\cite{Bourjaily:2011hi,Bourjaily:2015bpz}}. In \mbox{\cite{Bourjaily:2016evz}} a set of graphical rules (exploiting the correspondence between this amplitude and correlation functions) was described and used to determine the amplitude through ten loops; more recently, a new graphical rule was described by \cite{He:2024cej} which brought this benchmark to eleven loops! 

Although the only `observable' associated with this amplitude is the cusp anomalous dimension (well known to high orders and at strong coupling via integrability \mbox{\cite{Gromov:2009tv,Beisert:2010jr}}), studying this object at the \emph{integrand}-level has led directly to developments of generalized unitarity \mbox{\cite{Mangano:1990by,Bern:1994zx,Bern:1994cg,Dixon:1996wi,Cachazo:2005ga,Bern:2007dw,Feng:2011np}}, on-shell recursion relations \mbox{\cite{BCF,BCFW,ArkaniHamed:2010kv}}, the discovery of dual conformal invariance \mbox{\cite{Drummond:2007cf,Alday:2007hr,Drummond:2008vq,Brandhuber:2008pf,Drummond:2009fd}} and Yangian invariance \cite{Drummond:2010km} of sYM, along with much more. Moreover, by taking higher-point light-like limits, it is known that this one particular function captures \emph{complete} perturbative information about \emph{all} $n$-point scattering amplitudes in sYM \mbox{\cite{Brandhuber:2007yx,Caron-Huot:2010ryg,Alday:2010zy,Eden:2010zz,Eden:2010ce,Eden:2011ku,Eden:2011yp,Eden:2011we,Eden:2012tu,Heslop:2022xgp}} (see also~\mbox{\cite{Gonzalez-Rey:1998wyj,Eden:1998hh,Eden:1999kh,Eden:2000mv,Bianchi:2000hn}}). 

In this work, we argue that the \emph{double-triangle} rule of \cite{He:2024cej} in fact implies the `square' and `pentagon' rules described in \cite{Bourjaily:2016evz}, and we use this together with the `triangle' rule of \cite{Bourjaily:2016evz} to determine the integrand of correlator/amplitude to 12 loops. This new graphical rule relates $\ell$-loop contributions to those at $(\ell{-}1)$ loops, fixing all contributions at $(\ell{-}1)$ loops completely as we confirm at $\ell=12$. A byproduct is an independent confirmation of the so-called Catalan conjecture for the {\it anti-prism} graphs which have the largest coefficients, $-42$ in the $12$-loop case. Moreover, our results provide evidence for a much more general conjecture, which predicts coefficients of certain infinite families of graphs as common generalizations of anti-prism graphs and $f$-graphs for the so-called {\it fishnet} integrals~\cite{Basso:2017jwq}. Without such high-loop empirical `data', it is hard to imagine such structure being anticipated or discovered.

\vspace{-14pt}\section{Summary of the Double-Triangle Rule}\vspace{-14pt}
Consider the connected four-point correlation function $\mathcal{G}\equivR\langle\mathcal{O}(x_1)\,\bar{\mathcal{O}}(x_2)\,\mathcal{O}(x_3)\,\bar{\mathcal{O}}(x_4)\rangle$ of the lightest half-BPS operators $\mathcal{O}(x)\equivR\mathrm{tr}(\phi(x)^2)$ in sYM. Perturbatively, loop corrections of $\mathcal{G}$ can be computed using Lagrangian insertions~\cite{Eden:2011yp}: the $\ell$-loop \emph{integrand} is given by the \emph{Born-level} correlator with $\ell$ chiral Lagrangian insertions $x_{i=5,\ldots, 4{+}\ell}$. Normalized with an overall $\ell$-independent factor, the resulting \emph{integrand} is a rational function $\mathcal{F}^{(\ell)} (x_1, \ldots, x_{4{+}\ell})$ with uniform conformal weight ${-}4$ in all $(4{+}\ell)$ points enjoying complete $\mathfrak{S}_{4{+}\ell}$ permutation symmetry among its arguments~\cite{Eden:2011we}. Permutation invariance suggests that we describe this function in terms of unlabeled graphs, and expand $\mathcal{F}^{(\ell)}$ into a basis of $\ell$-loop `$f$-graphs' constructed as rational products of edge-factors. The space of inequivalent $f$-graphs can be easily constructed, allowing us to express $\mathcal{F}^{(\ell)}\equivL\sum_i c_i^{\ell}f_i^{(\ell)}$. 

Each $f$-graph corresponds to a permutation-invariant rational function constructed from a graph $\Gamma(f)$ involving $(4{+}\ell)$ vertices, with uniform conformal weight ${-}4$ in each. Letting $\Gamma_{\!\!\!\mathfrak{n},\mathfrak{d}}(f)$ denote the subgraphs associated with the numerator and denominator, respectively, we define 
\eq{f\equivR
\frac{\prod_{i\bullet\!\!-\!\!\bullet j\in\Gamma_{\!\!\!\mathfrak{n}}(f)}x_{i\,j}^2}{\prod_{i\bullet\!\!-\!\!\bullet j\in\Gamma_{\!\!\!\mathfrak{d}}(f)}x_{i\,j}^2}{+}\left(\substack{\displaystyle \text{permutations }\mathfrak{S}_{4+\ell}\\\displaystyle\text{mod }\mathrm{Aut}(\Gamma(f))}\right).}
That is, we may consider $f$-graph corresponds to an unlabeled graph $\Gamma(f)$ with $(4{+}\ell)$ vertices, and solid (resp., dashed) edges representing denominators $\Gamma_{\!\!\!\mathfrak{d}}(f)$ (resp., numerators $\Gamma_{\!\!\!\mathfrak{n}}(f)$). For example, the onlt planar (referring to the denominator's subgraph) $f$-graph at $\ell=3$ is\\[-12pt]
\eq{\fwbox{0pt}{\hspace{-10pt}\fig{-20pt}{1}{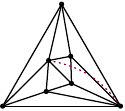}\bigger{\hspace{-5pt}\Leftrightarrow\;}\underbrace{\frac{x_{12}^2}{\prod_{i=3}^7x_{1i}^2x_{2i}^2x_{i,i+1}^2}\hspace{-2pt}+\hspace{-2pt}\left(\substack{\displaystyle\text{inequivalent}\\\displaystyle\text{perms.}}\right)}_{\text{252 distinct terms}},}}
which has $|\mathrm{Aut}(\Gamma(f))|=20$. We are interested in the planar limit of sYM, where only planar $f$-graphs contribute.

Universal divergences of the correlator under physical limits impose constraints which relate $\mathcal F^{(\ell)}$ and $\mathcal F^{(\ell-1)}$; they translate nicely into graphical rules which we use to bootstrap the $f$-graph coefficients. The most familiar example is the {\it triangle rule}~\cite{Bourjaily:2016evz} from the $\log x_{12}^2$ divergence in the OPE limit $x_2\to x_1$~\cite{Eden:2012tu}. A more powerful constraint was discovered in~\cite{He:2024cej} arising from the Sudakov $\log x_{12}^2\log x_{23}^2$ divergence in the double light-like limit $x_{12}^2,x_{23}^2\to0$, which leads to the {\it double-triangle rule}, $\mathcal P({\cal F}^{(\ell)})={\cal F}^{(\ell-1)}$ with the pinching operation $\mathcal P$ acting on all double-triangle structures:
\eq{\mathcal{P}\!\!:\fig{-19pt}{1}{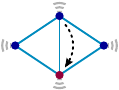}\bigger{\;\Rightarrow\;}\fig{-19pt}{1}{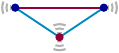}\,.\label{double_triangle_rule}}

\vspace{-14pt}\subsection{Square and Pentagon Rules are Implied by the Double-Triangle Rule}\vspace{-14pt}
The `square rule' of \cite{Bourjaily:2016evz} generalizes the `rung rule' of~\cite{Bern:1997nh} and can be derived from the consistency of the term $\mathcal{A}_1\mathcal{A}_{\ell{-}1}\subset{\mathcal{A}^2}_\ell$ arising from the perturbative expansion of the square of the amplitude (as determined from the correlator). As described in \cite{Bourjaily:2016evz}, it dictates equality between the coefficients of $f$-graphs that are related via
\eq{
\fig{-19pt}{1}{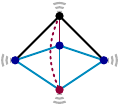}\bigger{\;\Leftrightarrow\;}\fig{-19pt}{1}{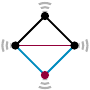}\,.\label{square_rule}}
It is easy to see this as a special case of the double-triangle rule, as the left-hand side is the unique pre-image of the resulting lower-loop graph under $\mathcal{P}$ (\ref{double_triangle_rule}), resulting in the necessary equality between the two coefficients. 

While powerful, there are many graphs not susceptible to the square rule. Taking inspiration from the square rule, but taking a five-point light-like limit resulted in the somewhat peculiar `pentagon' rule described in \cite{Bourjaily:2016evz} (to which we refer the reader for details). Essentially, the pentagon rule dictated that the sum of coefficients of a collection of $f$-graphs sharing a peculiar sub-topology must vanish. In light of the double-triangle rule, it is easy to now see that the pentagon rule also follows as a special case of the double-triangle rule:
\eq{
\left\{\fig{-24pt}{1}{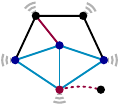}\,,\,\fig{-24pt}{1}{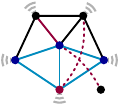}\right\}\bigger{\;\Rightarrow\;}\fig{-24pt}{1}{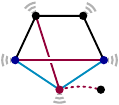}\,.\label{pentagon_rule}}
The collection of terms appearing on the left-hand side of (\ref{pentagon_rule}) are precisely those appearing as pre-images under $\mathcal{P}$ (\ref{double_triangle_rule}) of the non-planar graph on the right-hand-side. Because the image is non-planar, the double-triangle rule dictates that the sum of coefficients must vanish. 

\begin{table*}[t]\caption{Numbers of terms required to represent the $\ell$-loop amplitude via on-shell recursion, in terms of (dihedrally-symmetrized) dual-conformally invariant (`DCI') master integrals, or $f$-graphs---and how many have non-vanishing coefficients.}\vspace{-10pt}\label{contribution_statistics_table}$$\fwbox{00pt}{\text{\scalebox{0.9125}{$\hspace{0pt}\fwbox{0pt}{\begin{array}{rr@{$\,$}|@{$\,$}r@{$\,$}|@{$\,$}r@{$\,$}|@{$\,$}r@{$\,$}|@{$\,$}r@{$\,$}|@{$\,$}r@{$\,$}|@{$\,$}r@{$\,$}|@{$\,$}r@{$\,$}|@{$\,$}r@{$\,$}|@{$\,$}r@{$\,$}|@{$\,$}r@{$\,$}|@{$\,$}r@{$\,$}|}\\[-20pt]\multicolumn{1}{r}{\fwboxR{15pt}{\ell\,{=}\hspace{-1pt}}}&\multicolumn{1}{c}{1}&\multicolumn{1}{c}{2}&\multicolumn{1}{c}{3}&\multicolumn{1}{c}{4}&\multicolumn{1}{c}{5}&\multicolumn{1}{c}{6}&\multicolumn{1}{c}{7}&\multicolumn{1}{c}{8}&\multicolumn{1}{c}{9}&\multicolumn{1}{c}{10}&\multicolumn{1}{c}{11}&\multicolumn{1}{c}{12}
\\
\cline{1-13}\multicolumn{1}{|r|}{\fwboxR{53pt}{\text{recursed cells:\!}}}&\rule{0pt}{9pt}1&10&146&2\hspace{-0.5pt},\hspace{-1.5pt}684&56\hspace{-0.5pt},\hspace{-1.5pt}914&1\hspace{-0.5pt},\hspace{-1.5pt}329\hspace{-0.5pt},\hspace{-1.5pt}324&33\hspace{-0.5pt},\hspace{-1.5pt}291\hspace{-0.5pt},\hspace{-1.5pt}164&878\hspace{-0.5pt},\hspace{-1.5pt}836\hspace{-0.5pt},\hspace{-1.5pt}728&24\hspace{-0.5pt},\hspace{-1.5pt}175\hspace{-0.5pt},\hspace{-1.5pt}924\hspace{-0.5pt},\hspace{-1.5pt}094&687\hspace{-0.5pt},\hspace{-1.5pt}444\hspace{-0.5pt},\hspace{-1.5pt}432\hspace{-0.5pt},\hspace{-1.5pt}396&20\hspace{-0.5pt},\hspace{-1.5pt}086\hspace{-0.5pt},\hspace{-1.5pt}271\hspace{-0.5pt},\hspace{-1.5pt}785\hspace{-0.5pt},\hspace{-1.5pt}340&600\hspace{-0.5pt},\hspace{-1.5pt}384\hspace{-0.5pt},\hspace{-1.5pt}612\hspace{-0.5pt},\hspace{-1.5pt}445\hspace{-0.5pt},\hspace{-1.5pt}304\\
\cline{1-13}
\multicolumn{1}{|r|}{\fwboxR{14pt}{\text{DCI integrals:\!}}}&\rule{0pt}{9pt}1&1&2&8&34&278&3\hspace{-0.5pt},\hspace{-1.5pt}125&49\hspace{-0.5pt},\hspace{-1.5pt}935&981\hspace{-0.5pt},\hspace{-1.5pt}984&23\hspace{-0.5pt},\hspace{-1.5pt}045\hspace{-0.5pt},\hspace{-1.5pt}474&623\hspace{-0.5pt},\hspace{-1.5pt}496\hspace{-0.5pt},\hspace{-1.5pt}933&19\hspace{-0.5pt},\hspace{-1.5pt}117\hspace{-0.5pt},\hspace{-1.5pt}648\hspace{-0.5pt},\hspace{-1.5pt}284\\
\multicolumn{1}{|r|}{\fwboxR{14pt}{\text{{\small (\hspace{-1pt}contributing:\fwboxL{0pt}{\hspace{-2pt})}}\!}}}&\rule{0pt}{9pt}1&1&2&8&34&224&1\hspace{-0.5pt},\hspace{-1.5pt}818&19\hspace{-0.5pt},\hspace{-1.5pt}198&236\hspace{-0.5pt},\hspace{-1.5pt}823&3\hspace{-0.5pt},\hspace{-1.5pt}412\hspace{-0.5pt},\hspace{-1.5pt}129&56\hspace{-0.5pt},\hspace{-1.5pt}145\hspace{-0.5pt},\hspace{-1.5pt}999&1\hspace{-0.5pt},\hspace{-1.5pt}049\hspace{-0.5pt},\hspace{-1.5pt}691\hspace{-0.5pt},\hspace{-1.5pt}130\\
\cline{1-13}
\multicolumn{1}{|r|}{\fwboxR{14pt}{\text{$f$-graphs:}\!}}&\rule{0pt}{9pt}1&1&1&3&7&36&220&2\hspace{-0.5pt},\hspace{-1.5pt}707&42\hspace{-0.5pt},\hspace{-1.5pt}979&898\hspace{-0.5pt},\hspace{-1.5pt}353&22\hspace{-0.5pt},\hspace{-1.5pt}024\hspace{-0.5pt},\hspace{-1.5pt}902&619\hspace{-0.5pt},\hspace{-1.5pt}981\hspace{-0.5pt},\hspace{-1.5pt}403\\
\multicolumn{1}{|r|}{\fwboxR{14pt}{\text{{\small (\hspace{-1pt}contributing:\fwboxL{0pt}{\hspace{-2pt})}}\!}}}&\rule{0pt}{9pt}1&1&1&3&7&26&127&1\hspace{-0.5pt},\hspace{-1.5pt}060&10\hspace{-0.5pt},\hspace{-1.5pt}525&136\hspace{-0.5pt},\hspace{-1.5pt}433&2\hspace{-0.5pt},\hspace{-1.5pt}048\hspace{-0.5pt},\hspace{-1.5pt}262&35\hspace{-0.5pt},\hspace{-1.5pt}503\hspace{-0.5pt},\hspace{-1.5pt}735\\\cline{1-13}
\end{array}}$}}}\vspace{-18pt}$$\end{table*}

\vspace{-14pt}\section{Determining the 12-loop Correlator/Amplitude}\vspace{-14pt}
The computation of the 12-loop amplitude and correlator via the graphical bootstrap proceeded in four steps: (1) generating all $12$-loop $f$-graphs; (2) imposing the double-triangle rule to obtain bootstrap equations; (3) solving the bootstrap equations to obtain all visible 12-loop coefficients; and (4) using the triangle rule to fix the very few (three) remaining coefficients.

To generate all $f$-graphs at 12 loops, we started with all possible denominators, or planar graphs with 16 vertices and minimal valency 4, which was generated using \texttt{plantri}~\cite{Brinkmann2007FastGO} ignoring different embeddings. From these, it is straightforward to find all possible numerator `decorations' which give a valid $f$-graph (one of conformal weight $({-}4)$ in every vertex). Some planar graphs admit many numerators: at 12 loops, there is one admitting $213,\!082$ graphically-distinct numerators.

In order to impose the double-triangle rule relating the $\ell$-loop and $(\ell{-}1)$-loop coefficients, we first identify all possible ways of highlighting a double-triangle subgraph in an $\ell$-loop $f$-graph $f^{(\ell)}_i$. The pinching operation takes the double-triangle-highlighted $f$-graph $f^{(\ell)\triangleleft\hspace{-0.17ex}\triangleright}_i$ to a cusp-highlighted $f$-graph with one fewer vertex, $f^{(\ell-1)\vee}$. Including symmetry factors, the sum of pre-images for each cusp-highlighted, pinched graph must equal the coefficient of the lower-loop graph (zero if the pinched graph is not a planar or has double-poles). It is worth noting that applying these rules to each graph is easily parallelized: the double-triangle rule can be applied to one graph at a time. This was done using the high-performance computing cluster of ITP-CAS, requiring approximately $22,\!200$ core-hours to complete, resulting in $10,\!315,\!532,\!348$ bootstrap equations (with many duplications). Although the number of equations is large, they are extremely sparse, and easy to solve sequentially. It required 3 days to solve the resulting equations (albeit at a cost of hundreds of gigabytes of local memory). 

\begin{figure}\vspace{-14pt}$$\fig{-20pt}{1}{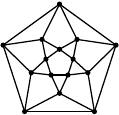}\qquad\fig{-20pt}{1}{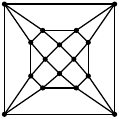}\qquad\fig{-20pt}{1}{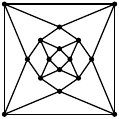}\vspace{-14pt}$$\caption{All 11,12-loop $f$-graphs without double-triangles---and hence `invisible' to double-triangle rule at the corresponding loop order. Interestingly, the 11-loop graph's coefficient \emph{is} determined by the $12{\to}11$-loop double-triangle rule.
}\label{fig:invisible}\end{figure}

\begin{figure}\vspace{-14pt}$$\fig{-20pt}{1}{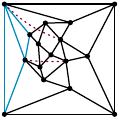}\qquad\fig{-20pt}{1}{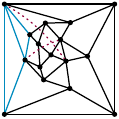}\vspace{-14pt}$$\caption{12-loop $f$-graphs whose coefficients are related \emph{but not determined} by the $12{\to}11$-loop double-triangle rule.
}\label{fig:nextto}\vspace{-16pt}\end{figure}

The resulting equations suffice to completely determine the 11-loop correlator/amplitude, and leave merely 3 undetermined 12-loop coefficients among the $619\hspace{-0.5pt},\hspace{-1.5pt}981\hspace{-0.5pt},\hspace{-1.5pt}403$ 12-loop $f$-graphs; these remaining coefficients were determined by application of the triangle rule. Specifically, in~\cite{He:2024cej} it was observed that the $\ell\!\to\!(\ell{-}1)$-loop bootstrap equations appeared to fully determine both all $(\ell\!-\!1)$-loop coefficients as well as all \emph{visible} $\ell$-loop coefficients---those with double-triangle sub-topologies among their denominators. Only one 11-loop graph is invisible, and only 2 at 12 loops; these are shown in \mbox{Fig.~\ref{fig:invisible}}. Although we confirm that the $12{\to}11$-loop bootstrap equations suffice to fully determine all 11-loop coefficients (including the `invisible' graph's), these equations did \emph{not} fully determine all `visible' 12-loop graphs' coefficients. The one exception is a pair of `next-to-invisible' graphs whose coefficients are related (but not fixed) by the double-triangle rule. These are shown in \mbox{Fig.~\ref{fig:nextto}}. These 3 coefficients were subsequently determined using the \mbox{(single-)}triangle rule.

It is interesting to compare these results to other possible methods. Using on-shell recursion~\mbox{\cite{ArkaniHamed:2010kv,ArkaniHamed:book,Bourjaily:2023apy,Bourjaily:2023uln}}, the 12-loop amplitude would require $>\!6\!\times\!10^{15}$ individual expressions~\footnote{Many of the on-shell diagrams counted in \mbox{Table \ref{contribution_statistics_table}} will vanish upon Fermionic integration. However, the number which vanish depends strongly on recursive choices made, but always appear to leave an $\mathcal{O}(1)$ fraction non-vanishing.}; and using unitarity with a basis of (dihedrally-symmetrized,) dual-conformally invariant master integrals, there would be $19\hspace{-0.5pt},\hspace{-1.5pt}117\hspace{-0.5pt},\hspace{-1.5pt}648\hspace{-0.5pt},\hspace{-1.5pt}284$ coefficients to determine. These are summarized in \mbox{Table~\ref{contribution_statistics_table}}.

Attached to this work's page on \texttt{arXiv}, the reader can obtain an ancillary file that includes \emph{all} contributions to the amplitude and correlation function through $\ell{=}12$ loops~\footnote{The reader may also download the data files from \url{https://huggingface.co/datasets/shicanxin/correlator_12loops/tree/main}}.

\vspace{-14pt}\subsection{Statistics of Coefficients}\vspace{-14pt}
We have given the statistics of $f$-graph coefficients up to $\ell{=}12$ in \mbox{Table~\ref{table:stats}}. These coefficients are in line with previous conjectures and observations. For example, half-integer coefficients appear for $\ell\!\geq\!8$, multiples of $\frac{1}{4}$ at $\ell\!\geq\!10$, and multiples of $\frac{1}{8}$ for $\ell\!\geq\!12$; and the coefficients of anti-prism graphs are given by (signed) Catalan numbers at even loop-orders.

Moreover, we identify interesting patters about the distribution of coefficients far from apparent at low loop orders. For example, the number of non-vanishing coefficients decreases rapidly with $\ell$: for $\ell{=}8,\ldots,12$, only about $39\%, 24\%,15\%,9.3\%,$ and $5.7\%$ of coefficients are non-zero. Moreover, most coefficients are concentrated within a very narrow ranges: \textit{e.g.}~for $\ell{=}10,11,12$ only $0.1\%, 0.04\%$ and $0.02\%$ lie beyond the range $[{-}1,1]$; at 12 loops, only $4\!\times\!10^{{-}6}$ fall outside $[-2,2]$, and a mere $2\!\times\!10^{{-}8}$ outside $[{-}5,5]$. From the data, it is clear that these ranges are dictated by Catalan numbers. Finally, we note the absence of any new coefficients between two consecutive Catalan numbers: for $\ell{=}8,9$ no coefficient lies within $({-}2,{-}5)$, for $\ell{=}10,11$ none between $(5,14)$, and for $\ell{=}12$ none within $({-}14,{-}42)$.

\begin{table*}\caption{Numbers of $f$-graphs contributing with each distinct coefficient for $\ell\!\leq\!12$. The vertical line is not to scale, but correctly ordered so as to highlight how new coefficients arise between gaps separating lower-loop coefficients.}\label{table:stats}\vspace{-14pt}$$%
\fwbox{0pt}{\hspace{-20pt}\fig{-40pt}{1}{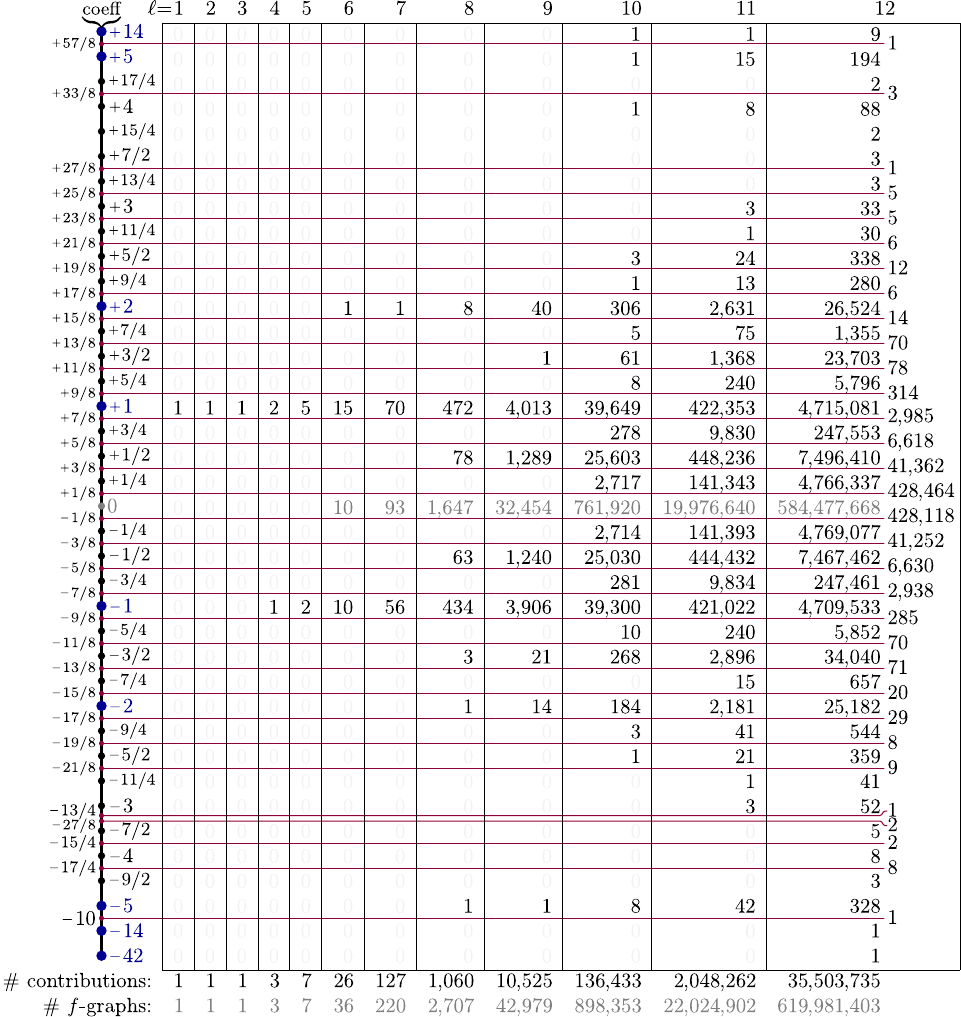}}\vspace{-14pt}$$\vspace{-04pt}\end{table*}

\begin{figure}\vspace{-14pt}$$\fig{-10pt}{1}{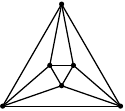}\fig{-10pt}{1}{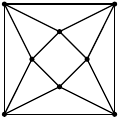}\fig{-10pt}{1}{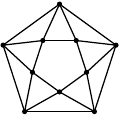}\fig{-10pt}{1}{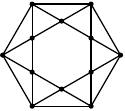}\fwboxL{0pt}{\hspace{-0pt}\ldots}\vspace{-12pt}$$\caption{Anti-prism graphs for $m\!\in\!\{3,4,5,6\}$.}\label{fig:antiprisms}\vspace{-14pt}\end{figure}

\vspace{-16pt}\section{The Generalized Catalan Conjecture}\vspace{-14pt}%
As we accumulate more and more data, patterns start to emerge that are wholly invisible at low loop orders. The most striking example is the Catalan conjecture proposed in~\cite{Bourjaily:2016evz}: at $2m$ points ({\it i.e.} $(2m{-}4)$ loops), the largest coefficient (in magnitude) is 
\eq{A_m\equivR(\mi1)^{m{-}1}C_{m{-}3}=(\mi1)^{m-1}\frac{1}{m{-}2}\binom{2(m{-}3)}{m{-}3}\,\label{calatan_numbers_defined}}
where $C_n$ denotes the $n$th Catalan number. For example, $A_m=1,{-}1,2,{-}5,14,{-}42$ for $m=3,\ldots,8$ (that is, for $\ell{=}2,4,6,8,10,12$). The corresponding $f$-graph with coefficient $A_m$ is the $2m$-point `anti-prism' illustrated in Fig.~\ref{fig:antiprisms}. The conjecture based on $m\!\leq\!7$ data was later confirmed at $m=8$ in~\cite{He:2024cej} using a {\it local system}. Our full calculation at $12$ loops confirms this completely.

\begin{figure}\vspace{-14pt}$$%
\fig{-20pt}{1}{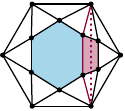}\;{\text{,}}\;\fig{-20pt}{1}{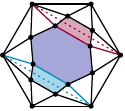}\quad\begin{array}{@{}c@{}}\text{not}\end{array}\quad\fig{-20pt}{1}{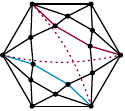}\vspace{-14pt}$$\caption{Two examples of valid, hexagon-framed fishnets, and one with an inconsistent numerator prescription; the coefficient of the last graph is zero.}\label{fig:threading}\end{figure}

\begin{figure}\vspace{-10pt}$$
\begin{array}{ccc}\begin{array}{@{}c@{}}\fig{-0pt}{1}{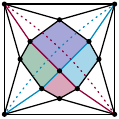}\\[-3pt]
\g{A_4}\b{A_4}\t{A_4}\r{A_4}
\end{array}&
\begin{array}{@{}c@{}}\fig{-0pt}{1}{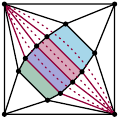}\\[-3pt]
\g{A_4}\b{A_4}\r{A_4}\t{A_4}
\end{array}&
\begin{array}{@{}c@{}}\fig{-0pt}{1}{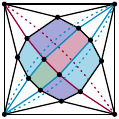}\\[-3pt]
\t{A_4}\b{A_4}\r{A_4}\t{A_4}\b{A_4}\g{A_4}
\end{array}\\
\begin{array}{@{}c@{}}\fig{-0pt}{1}{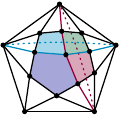}\\[-3pt]
\b{A_5}\t{A_4}\g{A_4}\r{A_4}
\end{array}&
\begin{array}{@{}c@{}}\fig{-0pt}{1}{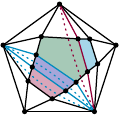}\\[-3pt]
\r{A_4}\b{A_4}\g{A_5}\t{A_4}
\end{array}&
\begin{array}{@{}c@{}}\fig{-0pt}{1}{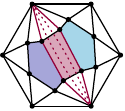}\\[-3pt]
\b{A_5}\r{A_4}\t{A_5}
\end{array}\\
\begin{array}{@{}c@{}}\fig{-0pt}{1}{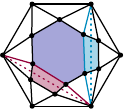}\\[-3pt]
\r{A_4}\b{A_6}\t{A_4}
\end{array}&
\begin{array}{@{}c@{}}\fig{-0pt}{1}{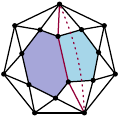}\\[-3pt]
\b{A_6}\t{A_5}
\end{array}&
\begin{array}{@{}c@{}}\fig{-0pt}{1}{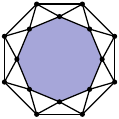}\\[-3pt]
\b{A_8}
\end{array}
\end{array}\vspace{-14pt}$$\caption{Several non-trivial examples of graphs with coefficients given by the generalized Catalan conjecture.}\label{fig:framed}\vspace{-14pt}\end{figure}

Now, with the complete data for $\ell\!\leq\!12$ at hand, we observe that this pattern generalizes to a broader class of $f$-graphs we denote as `polygon-framed fishnets'. These can be obtained as follows: start with a $2m$-point anti-prism, and repeatedly `thread' along diagonal directions~\footnote{More precisely, we require any two intersecting threads to intersect normally (not tangentially) at a valency-4 vertex, and all threads along the same diagonal direction must lie in parallel.
}. A `thread' is a sequence of denominators along a diagonal direction of the $m$-gon, together with a numerator connecting the endpoints of the diagonal~\footnote{Put differently, a polygon-framed fishnet is obtained by choosing a set of diagonals (including the $(i,i{+}2)$ diagonals already present) of a $2m$-point anti-prism, and thickening each diagonal into a bunch of threads.}. For example, start from the $m{=}6$ anti-prism of Fig.~\ref{fig:antiprisms}, sequentially threading along diagonals leads to the hexagon-framed fishnets such as the first two of Fig.~\ref{fig:threading}. Note that the third graph of Fig.~\ref{fig:threading} is \emph{not} a hexagon-framed fishnet due to the wrong placement of the numerators (although it is a valid $f$-graph). The polygon-framed fishnets are so named because the square-framed fishnets ($m{=}4$) (see \emph{e.g.}~the first three figures of Fig.~\ref{fig:framed}) are the $f$-graph versions (after dividing by $\xi_4$) of the more familiar rectangular fishnet integrals contributing to ${\cal G}$~\cite{Basso:2017jwq}.

The interesting observation is that, at least up to 12 loops, the coefficient of any polygon-framed fishnet is given by the product of $A_p$'s for each of its `faces' (each $p$-gon tile) excluding the outer polygon frame. 
%
For example, the $2m$-point anti-prism has a bunch of triangle tiles and an $m$-gon tile:
\eq{A_m \times \prod A_3=A_m\,.}
Any rectangular fishnet graph will have coefficient $\pm 1$, as the product consists of $A_3{=}1$ and $A_4{=}{-}1$ tiles; for a fishnet with $\ell{=}a\!\times\!b$ loops, its $f$-graph contains $(a{-}1)(b{-}1)$ squares, thus the coefficient is $({-}1)^{(a{-}1)(b{-}1)}$~\mbox{\cite{Caron-Huot:2021usw,He:2025vqt}}. 

\begin{figure}\vspace{-14pt}$$
\begin{array}{ccc}\begin{array}{@{}c@{}}\fig{-0pt}{1}{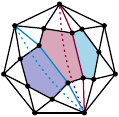}\\[-2pt]
\b{A_5}\r{A_5}\t{A_5}
\end{array}&
\begin{array}{@{}c@{}}\fig{-0pt}{1}{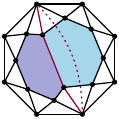}\\[-2pt]
\b{A_6}\t{A_6}
\end{array}&
\begin{array}{@{}c@{}}\fig{-0pt}{1}{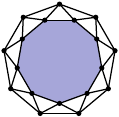}\\[-2pt]
\b{A_9}
\end{array}
\end{array}\vspace{-14pt}$$\caption{We predict the coefficients of these contributions to the 14-loop correlator to be $\b{A_5}\r{A_5}\t{A_5}=8$, $\b{A_6}\t{A_6}=25$, and $\b{A_9}=132$, respectively.}\label{fig:18pt:}\end{figure}

\begin{figure}\vspace{-16pt}$$%
\fig{-40pt}{1}{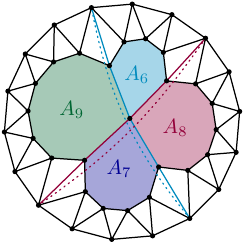}\vspace{-14pt}$$\caption{We predict the coefficient of this 37-loop graph to be: $\t{A_6}\r{A_8}\b{A_7}\g{A_9}=388,\!080$.}\label{37loop_prediction}\vspace{-14pt}\end{figure}

Polygon-framed fishnets are common generalizations of these two infinite families. See Fig.~\ref{fig:framed} for several non-trivial examples up to $12$ loops (with $m=5,6,7,8$). In particular, the final graph of  Fig.~\ref{fig:framed} is the 12-loop antiprism with coefficient $A_8{=}{-}42$ and the penultimate example has coefficient $A_5A_6{=}2\!\times\!({-}5){=}{-}10$. We conjecture this to hold to all loops: {\it e.g.} Fig.~\ref{fig:18pt:} shows predictions at $14$ loops, while Fig.~\ref{37loop_prediction} shows a prediction at $37$ loops.

While a proof of the original Catalan conjecture remains elusive, this evidence for its broader generalization strengthens our confidence that it will hold to arbitrary loop-orders; promisingly, as we describe in the Supplemental Material to this work, the coefficients appearing in the bootstrap equation involving the anti-prism also follow a nice pattern that resonates with a highly non-trivial identity of Catalan numbers.

\vspace{-18pt}\section{Outlook}\vspace{-14pt}
In this paper, we bootstrapped the 12-loop integrand of the four-point half-BPS correlator in planar sYM, using constraints from the leading divergent behavior of cusp and OPE limits. With enough computational resources, we are quite optimistic that such a bootstrap program could pursue several loop-orders higher. Meanwhile, a few more interesting questions naturally emerge.

First, it would be nice to explore whether these limiting behaviors completely characterize the four-point correlator, at least perturbatively, by proving whether these constraints (or the double-triangle rule by itself) suffice to determine the integrand to all loop orders.

Second, the integrand of the correlator provides valuable data for scattering amplitudes~\mbox{\cite{Ambrosio:2013pba,Heslop:2018zut}} and IR-safe weighted cross sections~\mbox{\cite{Hofman:2008ar,Belitsky:2013ofa,Henn:2019gkr,Yan:2022cye,Chicherin:2024ifn,He:2024hbb}} in sYM. Hopefully, the duality to amplitudes will provide a different perspective for the Catalan conjecture.

Third, it would be very interesting to understand the relation between the manifestly symmetric and local \mbox{$f$-graph} representation of the correlator and the more geometrical twistor representation~\mbox{\cite{Chicherin:2014uca,Eden:2017fow,He:2024xed}}.

\begin{figure}\vspace{-5pt}$$%
\fig{-20pt}{1}{anti10}\bigger{\;\Rightarrow\;}\fig{-20pt}{1}{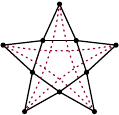}\vspace{-14pt}$$\caption{
A $5$-point integral seen as a pentagon-framed fishnet.}\label{fig:5ptintegral}\vspace{-14pt}\end{figure}

Another important direction is to study the higher-point Feynman integrals that these $f$-graphs provide (see \emph{e.g.}~Fig.~\ref{fig:5ptintegral}). It is likely that the polygon-framed fishnets, like the rectangular fishnets~\mbox{\cite{Gurdogan:2015csr,Caetano:2016ydc,Chicherin:2017cns,Basso:2017jwq}}, could be studied using integrability techniques~\mbox{\cite{Olivucci:2021cfy,Olivucci:2023tnw,Olivucci:2021pss,Olivucci:2022aza,Aprile:2023gnh}}.

Last but not least, the cusp limit could be applied more generally to nonplanar corrections~\cite{Fleury:2019ydf}, higher-point correlators~\mbox{\cite{Chicherin:2015bza,Bargheer:2022sfd}}, or heavier half-BPS correlators~\mbox{\cite{Chicherin:2015edu,Chicherin:2018avq,Caron-Huot:2021usw,Caron-Huot:2023wdh}}. In particular, since the generating function of heavier half-BPS correlators is represented by $f$-graphs lifted to 10 dimensions~\mbox{\cite{Caron-Huot:2021usw,Caron-Huot:2023wdh}} and thus also satisfies the graphical rules, a natural question is to explore the significance of these graphical rules in the 10d context. Could there be a cusped ``10d Wilson-loop'' that participates in a correlator generating function/``10d Wilson-loop''/Coulomb-branch amplitude triality?

\vspace{-14pt}\section*{Acknowledgments}\vspace{-10pt}%
\noindent It is our pleasure to thank Yao-Qi Zhang for fruitful discussions and initial collaboration on the work. The results described in this paper are supported by HPC Cluster of ITP-CAS. This work has been support by: a grant from the US Department of Energy \mbox{(No.\ DE-SC00019066)} (JB); the National Natural Science Foundation of China under Grant Nos.\ 12225510, 11935013, 12047503, 12247103; the New Cornerstone Science Foundation through the XPLORER PRIZE (SH); the China Postdoctoral Science Foundation under Grant No.\ 2022TQ0346 and the National Natural Science Foundation of China under Grant No.\ 12347146 (CS).

\begin{appendix}
\vspace{-14pt}\section{\texorpdfstring{Supplemental Material:\\Local System for Anti-Prism Graphs}{Supplemental Material: Local System for Anti-Prism Graphs}}
\begin{figure}$$\fig{-40pt}{1}{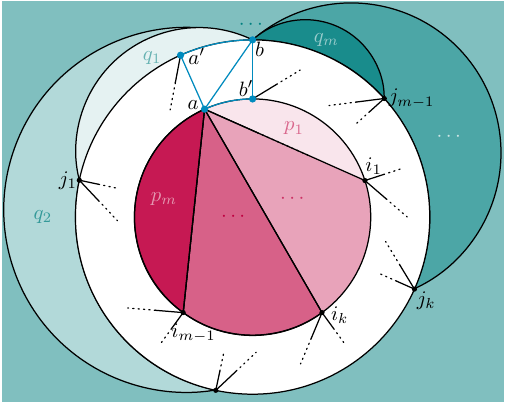}\vspace{-14pt}$$\caption{Typical diagram contributing to the bootstrap equation involving the $2n$-point anti-prism. Numerators $\prod_{k=1}^{m-1}x_{a,j_k}^2x_{b,i_k}^2$ are omitted for clarity. $\color{purple}p_k$/$\color{teal}q_k$ denote the sizes of faces colored purple/teal, respectively.}\label{fig:localsystem}\end{figure}

\vspace{-14pt}
Approaching the Catalan conjecture from a different perspective, we examine the unique bootstrap equation involving the $2n$-point anti-prism (corresponding to pinching the unique double-triangle subgraph on the ``belt''). The $f$-graphs involved in this equation (Fig.~\ref{fig:localsystem}) are characterized as follows. Suppose the highlighted double-triangle is $(\triangle aba')(\triangle abb')$. For some \mbox{$1\!\leq m\!\leq\!n-2$}, shoot $(m{-}1)$ ray-like diagonals from $a$ on the inner $n$-gon and another $(m{-}1)$ ray-like diagonals from $b$ on the outer $n$-gon, dividing the two $n$-gon faces into $2m$ faces of size $\color{purple}p_k$/$\color{teal}q_k$. Label the endpoints $i_1,\ldots,i_{m-1}$ of the inner diagonals \emph{clockwise} and the endpoints $j_1,\ldots,j_{m-1}$ of the outer diagonals \emph{counterclockwise}. These $2(m{-}1)$ denominators $(\prod_{k=1}^{m-1}x_{a,i_k}x_{b,j_k})^{-1}$ are compensated by a numerator $\prod_{k=1}^{m-1}x_{a,j_k}x_{b,i_k}$ to restore the correct conformal weight. These $f$-graphs pinch to the same cusp-highlighted $f$-graph as the $2n$-point anti-prism because the extra denominators and numerators precisely cancel.

We shall represent these $f$-graphs by the sequence $(p_1,\ldots,p_m;q_1,\ldots,q_m)$. It is easy to count that there are $\binom{2(n{-}3)}{n{-}3}$ allowed sequences by counting ways to shoot out the diagonals, but among these, the pair $(p_1,\ldots,p_m;q_1,\ldots,q_m)$ and $(q_1,\ldots,q_m;p_1,\ldots,p_m)$ lead to isomorphic $f$-graphs, since flipping one inside-out yields the other. Surprisingly, we find that the coefficients of these $f$-graphs are all described by the formula:
\eq{\begin{split}&c(p_1,p_2,\ldots,p_m; q_1,q_2,\ldots,q_m) = ({-}1)^{m{+}\!\sum_{k=1}^{m}p_k}\\
    &\times\prod_{k=1}^{m} C(p_{1,k}{-}q_{m-k+2,m}{-}3) C(q_{m-k+1,m}{-}p_{1,k}),\end{split}}
where $C(p)$ is the $p$th Catalan number for \mbox{$p\!\geq\!0$} and $C(p)\equivR0$ for $p\!<\!0$. We use the shorthand notation $p_{k,k'}\equivR p_k{+}p_{k+1}{+}\ldots{+}p_{k'}$ and similarly for $q_{k,k'}$. For example, the anti-prism coefficient is $c(n;n)=({-}1)^{n{+}1}C(n{-}3)$, which agrees with the Catalan conjecture. Other examples include $c(3,3,\ldots,3; 3,3,\ldots,3){=}{+}1$, which can be verified by recursively using the ``square rule'' to reduce to the unique planar 6-point $f$-graph. Another nontrivial check is that this formula is indeed invariant under exchanging $(p_1,\ldots,p_m)\!\leftrightarrow\!(q_1,\ldots,q_m)$. This formula successfully reproduces the correct coefficients for all such graphs up to $16$ points. Furthermore, we checked that up to $2n\leq28$ points, the $\binom{2(n{-}3)}{n{-}3}$ coefficients always sum up to 0, as they should according to the bootstrap equation.
\end{appendix}

\vspace{-14pt}
\providecommand{\href}[2]{#2}\begingroup\raggedright\endgroup

\end{document}